\documentclass[12pt, preprint]{aastex}
\usepackage{amsmath,amsthm}
\usepackage{color}
\usepackage{rotate}
\usepackage{rotating}

\newcommand{\E}{\mathcal{E}}

\newcommand{\G}{\mathcal{G}}
\newcommand{\M}{\mathcal{M}}
\newcommand{\F}{\mathcal{F}}

\def\arcsec{\hbox{$^{\prime\prime}$}}
\def\arcmin{\hbox{$^{\prime}$}}

\def\cc{cm$^{-3}$}
\def\kms{\ensuremath{{\rm{}km\,s^{-1}}}}

\def\htwo{H$_2$}

\def\ammonia{NH$_3$}

\def\n2h{N$_2$H$^+$}
\def\Ms{\ensuremath{{\rm{}M}_{\odot}}}
\def\13co{$^{13}$CO}
\def\c18o{C$^{18}$O}
\def\nh2{$n(H_2$)}

\def\h13co{H$^{13}$CO$^+$}
\def\micron{$\mu$m}
\def\lp{\>\> .}
\def\lc{\>\> ,}
\newcommand{\C}{\mathcal{C}}
\newcommand{\R}{\mathcal{R}}

\begin{document}

\title{MASSIVE QUIESCENT CORES IN ORION: DYNAMICAL STATE REVEALED BY HIGH-RESOLUTION AMMONIA MAPS}
\author{D.\  Li\altaffilmark{1,2,3}, J. Kauffmann\altaffilmark{4},
  Q. Zhang\altaffilmark{5}, W. Chen\altaffilmark{6}}
\altaffiltext{1} {National Astronomical Observatories, Chinese Academy
  of Science, Chaoyang District Datun Rd A20, Beijing, China;
  \emph{email:} ithaca.li@gmail.com}
\altaffiltext{2} {Space Science Institute, Boulder, CO, USA}
\altaffiltext{3} {Jet Propulsion Laboratory, California Institute of
  Technology, Pasadena, CA, USA}
\altaffiltext{4} {Department of Astronomy, California Institute of Technology, 1200 E California Boulevard, Pasadena, CA 91125, USA}
\altaffiltext{5} {Harvard-Smithsonian Center for Astrophysics,
  Cambridge, MA, USA}
\altaffiltext{6} {Peking University, Beijing, China}

\begin{abstract}
   We present combined VLA and Green Bank Telescope images of \ammonia\
  inversion transitions (1,1) and (2,2) toward OMC2 and OMC3. We
  focus on the relatively quiescent Orion cores, which are away from the
  Trapezium cluster and have no sign of massive protostars nor evolved
  star formation, such as IRAS source, water maser, and methanol
  maser.  The 5\arcsec\ angular resolution and $0.6~\rm{}km\,s^{-1}$ velocity
  resolution of these data enable us to study the thermal and dynamic
  state of these cores at $\sim{}0.02~\rm{}pc$ scales, comparable to
  or smaller than those of the current dust continuum surveys. We
  measure temperatures for a total of 30 cores,
  with average masses of $11\,\Ms$, radii of $0.039~\rm{}pc$, virial mass ratio $\overline{R_{vir}}$ = 3.9, and critical mass ratio $\overline{R_{C}}$ = 1.5.
  Twelve sources contain \textit{Spitzer} protostars. The thus defined starless and protostellar subsamples have
  similar temperature, line width, but different masses, with an average of $7.3\,\Ms$ for the former and $16\,\Ms$ for the
  latter. Compared to others Gould Belt dense cores, mores Orion cores have a high gravitational-to kinetic energy ratio
  and  more cores have a larger thant unity critical mass ratio. Orion dense cores have velocity dispersion similar to
  those of cores in low-mass star-forming regions but larger masses for fiven size. Some cores appear to have truly
  supercritical gravitational-to-kinetic energy ratios, even when considering significant observational uncertainties:
  thermal and non-thermal gas mothins alone cannot prevent collapse.
\end{abstract}

\section{INTRODUCTION}
Stars form in molecular clouds. Within molecular clouds, discrete
structures with observable column density contrast, particularly in
high density tracers, with its surroundings are referred to as cores (e.g.\
Benson \& Myers 1989; Ward-Thompson et al.\ 2007). The relatively high
extinction and density of cores make them the likely site for the
onset of collapse.

Recent core surveys (e.g.\ Ikeda et al.~ 2007; K{\"o}nyves et al.\ 2010; Sadavoy et al.\ 2010) are facilitated by focal plane imaging
arrays of growing sizes.  Although many surveys have large sample size
in the hundreds, the majority of these studies are of dust continuum, which tend to focus on core mass function. Direct
measurement of cores' thermal and dynamic structures require spectra
maps of high density tracers and preferably cover the same spatial
dynamic ranges as those of the dust emission.

There seems to be an observational dichotomy between low mass star formation and high
mass star formation.  Massive stars are formed exclusively in giant
molecular clouds and with higher efficiency.  Due to the large
distances of most massive star forming regions, many massive cores are
under-resolved, showing signs of star formation well underway, such as
H$_2$O masers (Mookerjea et al.~2004) and/or compact HII regions (Reid \& Wilson 2005).
At about 437 pc (Hirota et al.~2007; Menten et al.\
2007), Orion molecular clouds are the closest massive star forming
region with an OB cluster, thus particularly suited for studying the
early stages of star formation in massive cores and/or under the
influence of young massive stars.  In a series of papers, we
identified 'quiescent' Orion  clouds/cores (containing no HII region, no IRAS
point sources, and are at last 1 pc away from the OB cluster)
with \ammonia\ and \n2h\ with a beam size of about 1\arcmin\
(Li, Goldsmith \& Menten 2003; hereafter paper I), presented high
resolution 350 \micron\ images with a beam size of 9\arcsec\ (Li et
al.\ 2007; hereafter paper II), and revealed that two thirds of Orion
cores have signatures of ongoing dynamic evolution, both outflows and
inflows (Velusamy et al.\ 2008 hereafter paper III). Based on dust
mass and dust core size (paper II), the majority of the cores are
seemingly supercritical, i.e., no adequate support from thermal
pressure, turbulence(c.f. McKee \& Tan 2003), or static magnetic field.  This is consistent
with the majority of the cores being hydrostatically unstable (paper
III), which, however, could not be directly tested due to a lack of
high resolution spectroscopic data.

Spectroscopic survey of CO 1-0 (Williams, Plambeck \& Heyer 2003) and
CO 3-2 (Takahashi et al.~2008) reveal that part of the Orion
molecular cloud, namely, OMC2 and OMC3, contains many molecular
outflows.  OMC2 and OMC3 have also been mapped in high density
tracers. Using the Nobeyama 45m telescope, Tatematsu et al.\ (2008)
identifed 34 cloud cores in \n2h\ 1-0 with a beam width of around
18\arcsec. Ikeda et al.\ (2007) studied dense gas in the same region
using H$^{13}$CO$^+$ also with Nobeyama.  These studies find higher
column density for respective tracers compared with low mass star
forming regions, such as Taurus. \ammonia\ inversion transitions are
particularly suited for studying dense cores due to their lack of
depletion and their sensitivity to kinetic temperature (Ho \& Townes
1983).  Wiseman \& Ho (1998) obtained \ammonia\ maps of a 8\arcmin\
$\times$ 6\arcmin\ region around Orion-KL using VLA.  The gas
morphology there is dominated by quasi-parallel filaments severely
influenced by the energy output of young massive stars.

In this letter, we present the combined VLA and GBT ammonia survey of
the cores in OMC2 and OMC3.  The spatial resolution of $\sim$
5\arcsec\ and the spectral resolution of 0.6 \kms\ facilitate a
detailed examination of the thermal and dynamic states of massive
quiescent Orion cores.

\section{OBSERVATIONS AND DATA REDUCTION}
The VLA observations of OMC2/OMC3 were carried out in the D
configuration on July 29 and September 24 of 2000. We used the
correlator mode 4 to cover the \ammonia\ (J,K) = (1,1) and (2,2)
inversion transitions simultaneously. With a primary beam of about
2$'$ at the observing frequencies, a total of 20 pointings were used
to mosaic the OMC2/3 regions.  The correlator was configured to a
bandwidth of 3.13 MHz for each transition that was divided into 64
channels, providing a velocity coverage of $40~\kms$ in $0.6~\kms$
channel spacing. We used quasar 3C286 as the
flux calibrator and 3C84 or 3C273 as bandpass calibrators.
 The time dependent gains were monitored by observing
0539-057 or 0605-085. Visibility data were calibrated in AIPS and
exported to MIRIAD for imaging. The 1$\sigma$ RMS in the
channel maps, after combining the two observations is about
$8~\rm{}mJy$ per spectral channel.

The observations of \ammonia\ (1,1) and (2,2) transitions with the GBT
were taken on April 6th and 7th of 2005 in the OTF mode using the
receiver Rcvr18-26. The correlator setup had a 200 MHz bandwidth and
$0.31~\kms$ velocity resolution (8192 channels).  The flat baseline and
the wide bandwidth allowed us to use frequency switch with a throw of
12.5 MHz. The spectra were folded and the 3D data cube were resampled
to a grid of 12$''$ spacing and $0.3~\kms$ channel. The resulting GBT
spectra has a typical $\rm{}0.12~\rm{}K$ RMS noise per channel.

\begin{figure*}[htp]
\includegraphics[width=1.0\textwidth,angle=0]{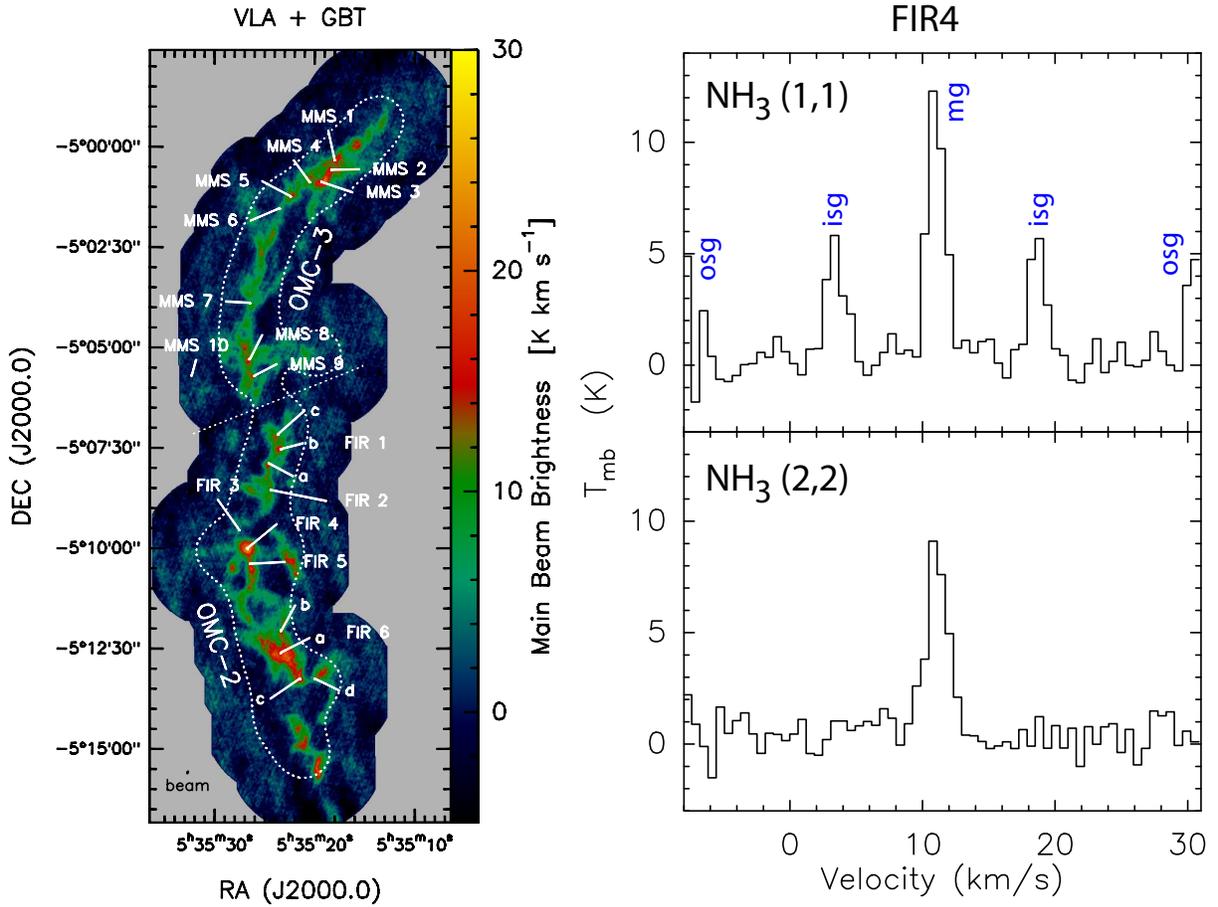}
%\vspace{5cm}
\caption{The integrated intensity of \ammonia\ (1,1) overlaid with
  dust continuum peaks from Chini et al.\ (1997) and references
  therein. The white dotted line indicates the location where the gain
  drops to 80\% of the peak value.  The spectra shown on the right are
  for the FIR 4 position, the brightest \ammonia\ peak in OMC 2. The VLA bandwidth covers
  fully the main group (mg) and the inner satellite groups (isg) of the hyperfine components, and only partially the outter satellite groups (osg).}
\label{nh3}
\end{figure*}

We combine the \ammonia\ data from the VLA with those from the GBT to
recover missing short spacing fluxes in the interferometric data. The
combination was performed in the UV domain using MIRIAD, following the
procedure outlined in Vogel et al. (1984) and Zhang et al. (2000). The
\ammonia\ emission from the combined image, when convolved to the
30$''$ GBT beam, recovers more than 80\% of the fluxes detected in the
GBT data.  The integrated intensity of the \ammonia\ (1,1) line with
previously identified continuum sources and a pair of typical spectra
are presented in Fig.~\ref{nh3}.

\section{Derivation of Kinetic Temperature and Velocity Dispersion}
Largely following Ho \& Townes (1983), Paper I described a recipe for
deriving kinetic temperature from spectrally resolved, modestly blended \ammonia\ lines (intrinsic line
width $\Delta{}V\sim{}1.0~\kms$). In this paper, we report line width
$\Delta{}V$ in FWHM. The one dimensional velocity dispersion $\sigma$
used in Eq.~\ref{mvir}, Eq.~\ref{mj} and Eq.~\ref{pic} is related to
FWHM as $\Delta{}V=\sqrt{8\ln(2)}\,\sigma$ for a Gaussian. The key
step is to obtain the optical depth of the (1,1) line from
simultaneously fitting all hyperfine components. As can be seen in
Fig.~\ref{nh3}, the VLA band is not wide enough to fully cover the two
outer groups ($osg$). The combined data set thus only have the main
and inner satellite components of the (1,1) transition and have a
velocity resolution of $0.6~\kms$. Toward most positions, the intrinsic
line width is smaller or approaching channel width.  For such
under-resolved line, the opacity cannot be easily uniquely fitted.

\begin{figure}[htp]
\begin{center}
\includegraphics[width=1.0\linewidth]{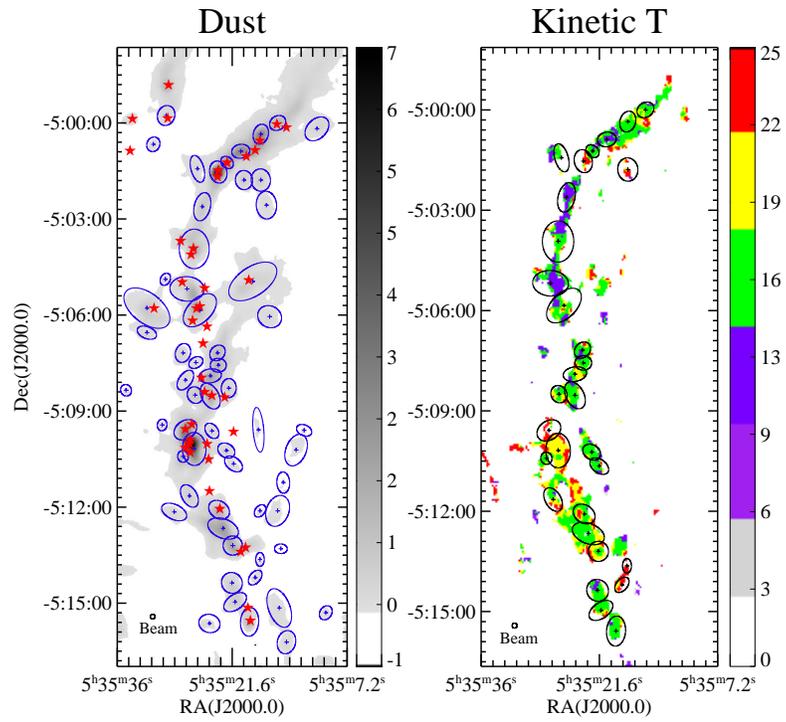}
%[width=200mm]{Dust+tk.eps}
\caption{Left: 850 \micron\ dust\ images with dust cores (N07).  The red stars are embedded YSOs (Megeath et al.~2012) Right: Derived kinetic temperatures overlaid with cores. }
\label{tk}
\end{center}
\end{figure}

We developed a more straightforward recipe utilizing two ratios
between integrated intensities, which are directly observable,
$\R^{12}=[\int{}T^{(mg+isg)}_A{}(1,1){}d\nu]/[\int{}T_A{}(2,2)d\nu]$
and
$\R^{sm}=[\int{}T^{(isg)}_A{}(1,1)d\nu]/[\int{}T^{(mg)}_A{}(1,1)d\nu]$.
The rotational temperature can be derived as the following
\begin{equation}
T_R=41.5~{\rm{}K}/\ln{}[1.06\times{}\C(1,1)\times{}R^{12}]\lc
\label{tr}
\end{equation}
where $\C(1,1)$ is a numerical factor  determined as
\begin{equation}
\C(1,1)=0.003+2.26\,\R^{sm}+0.00032\,{\rm{}e}^{5.38\,\R^{sm}}\lc
\label{cprime}
\end{equation}
which is based on fitting the simulated \ammonia\ (1,1) spectra
based on  a grid of opacities and linewidth.
The kinetic
temperature is then (paper I)
\begin{equation}
T_k=3.67+0.307 \times T_R+0.0357 \times T^2_R  \lp
\end{equation}
The full recipe is given in Kauffmann et al.\ (2012) and is generally
applicable to a wide range of conditions in opacities, channel width,
and intrinsic linewidth.  In the observed linewidth range, the dependence
of $\C(1,1)$ on linewidth is small. The uncertainty in derived kinetic
temperature is then calculated using a Monte Carlo approach (paper I).
For pixels with (2,2) detection, the representative 1$\sigma$
uncertainty is about $1~{\rm{}to}~2~\rm{}K$ for emission peaks and
between $2~\rm{}K$ and $5~\rm{}K$ for diffuse areas. Except for about
1.5\% of scattered pixels, the derived temperatures are between
$10~\rm{}K$ and $30~\rm{}K$.  When there is no detection of the (2,2)
line, we derive an upper limit to kinetic temperatures ($T^u_k$)
assuming 3$\sigma (2,2)$ intensity. Figure~\ref{tk} shows the derived
temperatures overlaid with the dust continuum and dust cores.  The
temperatures  generally drops while moving inward the dust cores,
consistent with paper I at a coarser spatial grid.  A quantitive analysis of the
thermal structure of the whole region will be presented in Kauffmann et
al.\ (2012).

The low velocity resolution of $0.62~\rm{}km\,s^{-1}$ complicates the
calculation of velocity dispersions. We first identify main group
channels with signal--to--noise ratios $>3$. If more
than one such channel is found, a dispersion $\sigma{}_{\rm{}data}$
is calculated, in which channels are weighted by their intensity; if a
single channel is found, an upper dispersion limit  of $0.26~\rm{}km\,s^{-1}$ (channel width) is
adopted .  The intrinsic
dispersion of the (1,1) main group is then subtracted to obtain
$\sigma_{\rm{}obs}=[\sigma_{\rm{}data}^2-(0.2~{\rm{}km\,s^{-1}})^2]^{1/2}$.
A numerical experiment was performed based on simulated Gaussian lines with noise.  For
our observed core-averaged SNR  of 13 to 43 and typical dispersion of $<$0.4 \kms,  using only channels with SNR$>$3 results in an underestimation of the intrinsic line dispersion  by about 5\%.
This is negligible considering that we have used an upper limit for
single channel detections.

To account for the internal centroid velocity variation, the standard deviation ($\sigma_{pp}$)
among the peak velocities of spectra contained within each core was added to the pixel averaged velocity dispersion ($\sigma_{obs}$) as  $\sigma_{core}^2 = \sigma_{obs}^2+\sigma_{pp}^2$.
The core velocity dispersion ($\sigma_{core}$) thus derived along with pixel-to-pixel velocity deviation
are  listed in Table 1.

\section{Dynamic State of Cores}
For a full coverage of all OMC2 and OMC3 cores, we use the SCUBA
survey of Orion by Nutter \& Ward-Thompson (2007, hereafter N07), who identified
cores as ellipses in the signal to noise ratio map.  They stated that the flux calibration is accurate to
the level of about 10\%.
The uncertainty in the determination of dust mass comes mainly from the assumption of dust opacity, which is $45\%$ lower in N07 than in Ossenkopf \& Henning (1994). This should be kept in mind as
a generic uncertainty in derived dust masses.
We used the published dust mass in  N07 and  implement the following
correction based on temperatures derived in this paper.

We select SCUBA
cores with more than 50\% of the pixels having a measured $T_k$ based
on \ammonia.  These temperature measurements are weighted by dust
column density and then averaged to be  the core
temperature $T_{core}$.   A uniform dust temperature of 20 K was assume in N07.
In our map, 74\% of the pixels have derived temperature lower than 20 K.  We
therefore scale the core mass by the factor
$B_{850,20~{\rm{}K}}/B_{850,T_{core}}$, where $B_{850,T}$ is the
Planck function at $850~\rm{}\mu{}m$ and temperature $T$.  The
resulting core mass is generally larger than what is reported in N07. The derived core temperatures and core
velocity dispersion are reported in Table~\ref{tab1}.

We identify cores with and without embedded stars based on Spitzer
source identifications (Figure 2). A core is deemed
protostellar if the distance between its peak position and a YSO are
less than the semi-minor axis of the core ellipse (Fig.~\ref{tk} and
Table~\ref{tab1}).

\subsection{Are Cores Gravitationally Bound?\label{sec:r-vir}}
McKee \& Zweibel (1992) derive the virial theorem in Eulerian form to
express explicitly the contribution of turbulent pressure both inside
and surrounding the cloud

\begin{equation}
2c_w(\E-\E_0)+\M_s+\G=0\lc
\label{vir}
\end{equation}
where
\begin{equation}
c_w=1+\frac{\E_w-\E_{0w}}{2(\E-\E_0)}
\end{equation}
is determined by the ratio of the difference between surface turbulent
energy $\E_{0w}$ and internal turbulent energy $\E_w$, and total
surface kinetic energy $\E_0$ and total internal kinetic energy
$\E=\frac{3}{2}M\sigma^2$.  For a cloud with primarily turbulent
motions, $c_w=3/2$. If steady motions, such as rotation, are
significant, $c_w$ is close to unity. The velocity resolution of our
data limits our ability to evaluate directly the role of turbulence, which could
be dominant (McKee \& Tan 2003).  For a significant fraction of the pixels, we
only have an upper limit to the linewidth, which results in an overestimation of the
total core velocity dispersion.  A unity $c_w$ is thus assumed
for simplicity for the virial mass calculation.
$\M_s$ is the magnetic energy associated with the cloud.  $\G$ is the
gravitational potential energy.  For an axisymmetric ellipsoid with
concentric density profiles
\begin{equation}
\G =-\frac{3}{5}\alpha{}\beta{}\frac{GM^2}{r}\lc
\end{equation}
where $\beta=\arcsin{e}/e$ is the geometry factor determined by
eccentricity $e=\sqrt{1-f^2}$ and $\alpha=(1-a/3)/(1-2a/5)$ for a
 power law density profile $\rho\propto{}r^{-a}$.
 We adopt $a=1.6$ for an isothermal cloud in equilibrium (Bonnor 1956).
The intrinsic axis ratio $f$ is smaller than the
observed axis ratio $f_{obs}$ due to projection. Myers et al.\ (1991)
show that dark cloud cores are more likely to be
prolate. Condensations in Orion are also statistically more consistent
with being prolate rather than oblate (Li 2002). Based on the more
general inversion from Fall \& Frenk (1983), we derive an averaged
true axis ratio $\bar{f}$ corresponding to a sample of a single value,
namely, the observed $f_{obs}$ for a prolate ellipsoid
\begin{equation}
\bar{f}=\frac{2}{\pi}f_{obs}\F_1(0.5,0.5,-0.5,1.5,1,1-f^2_{obs})\lc
\label{prolate}
\end{equation}
where $\F_1$ is the Appell hypergeometric function of the first
kind, which gives $\bar{f}=1$ for $f_{obs}=1$ and
$\bar{f}=0$ for $f_{obs}=0$. For values in between, we
 $\bar{f}$ as a replacement of $f$ in calculating the gravitational potential.

\begin{figure}[htp]
\includegraphics[width=1.0\linewidth]{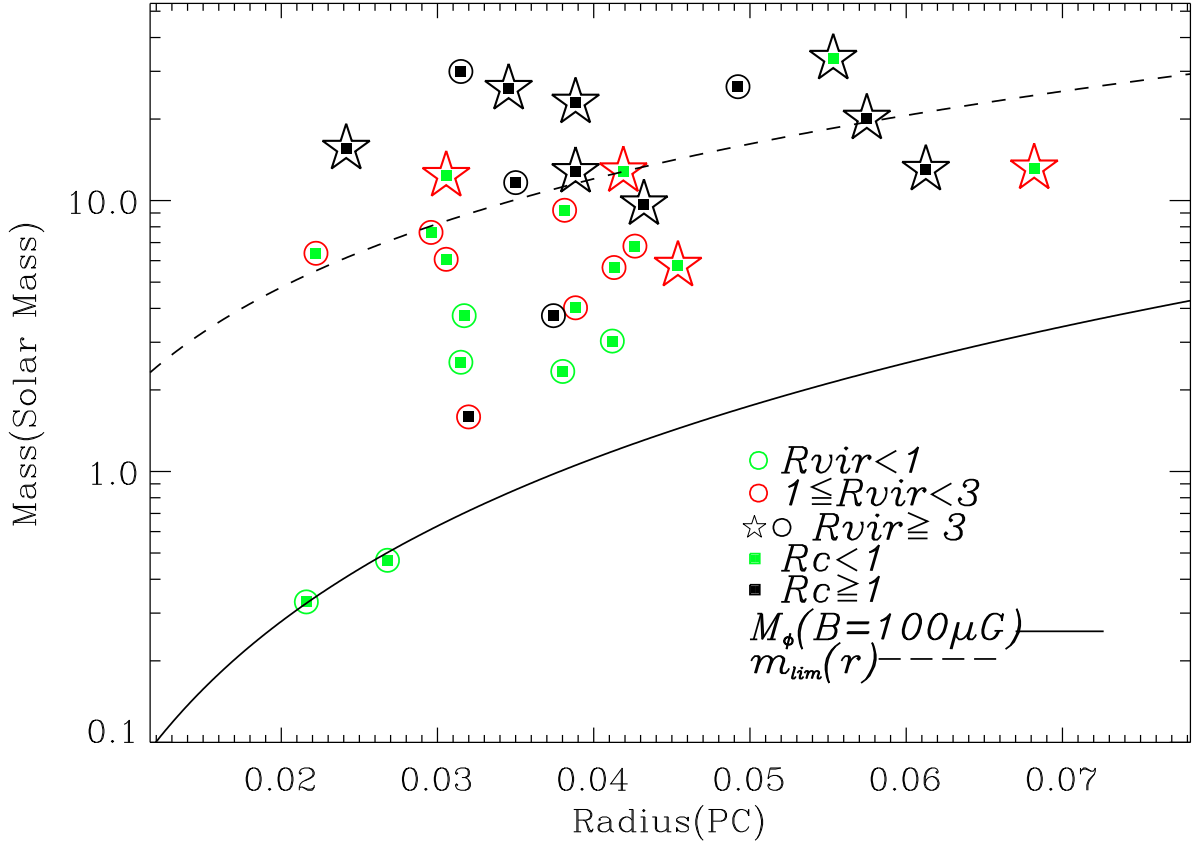}
\caption{The core mass-size relationship similar to that in paper II.  The colors of the open symbols denote the virial mass
  ratio $R_{vir}$, with green for $R_{vir}>1$,
  red for  $1<R_{vir}<3$ and black for $R_{vir}>3$. Open
  stellar symbols are protostellar cores and open
  circles are starless cores. The colors of the solid squares denote the critical mass
  ratio $R_c$, with green for $R_c<1$ and black for $R_c>1$. The solid curve is for mass
  supported by a steady magnetic field of $100~\rm{}\mu{}G$.
  The dashed curve represents the empirical threshold for massive star formation (Eq.~\ref{mlim}). }
\label{mass}
\end{figure}

By assuming $\M_s=0$ and $\E_0=0$, a simple virial mass
can be derived using the virial theorem (Eq.~\ref{vir})
\begin{equation}
M_{vir}=\frac{5}{\alpha{}\beta}\frac{\sigma^2{}r}{G}\lp
\label{mvir}
\end{equation}
The virial mass is the minimal mass for a cloud to be
self--gravitating in the absence of pressure confinement and steady
magnetic field. A "virial mass ratio" independent of $B$ and $\E_0$ is
defined as
\begin{equation}
R_{vir}=\frac{M}{M_{vir}}\lp
\label{rvir}
\end{equation}
We find that 24 of 30 cores  (Fig.~\ref{mass} and
  Table~\ref{tab1}), i.e., 80\% have $R_{vir}>1$, 11 cores (37\%) fulfil
  $R_{vir}>3$.  The majority of the core sample are bound by gravity.

\subsection{Stability and Critical Mass\label{sec:r-c}}
The stability of cores are considered here in the context of external pressure and
the internal magnetic field. The critical mass $M_C$ is
defined as the maximum mass which can be stably supported
by internal velocity dispersion and magnetic pressure.
The two effects can be considered separately and then
combined in a simple approximation
\begin{equation}
M_C=M_J+M_\Phi{}\lc
\label{mc}
\end{equation}
which is accurate to within 5\%
compared to the more rigorous calculations (McKee 1989).

The Jeans mass for a non-magnetic isothermal cloud (Bonnor 1956; McKee \& Zweibel 1992) is
\begin{equation}
M_J=1.182\frac{c_w^{3/2}}{c_{pr}^{1/2}}\frac{\sigma^4}{G^{3/2}P_{ic}^{1/2}}\lc
\label{mj}
\end{equation}
where $c_w$ parametrizes the pressure from turbulence outside of the
cloud and $c_{pr}$ parametrize the added internal pressure from
turbulence's disturbing B field.  Jeans mass depends on the fourth power of the
line width.  If we consider tracers
of more diffuse gas in this region, such as \13co, the linewidth is
around $1.5~\kms$ (Melnick et al.\ 2011), definitely supersonic.  We
choose the combined factor to be in the middle of the possible range
for turbulent gas $c_w^{3/2}/c_{pr}^{1/2}\sim{}2.1$ (McKee \& Zweibel 1992).
The external pressure term can be estimated as
\begin{equation}
P_{ic}=n_{ic}\mu{}m_H\sigma_{ic}^2{}\lc
\label{pic}
\end{equation}
where the mean molecular weight $\mu$ relative to n(\htwo)
is 2.8 (Kauffman et al.~2008) and $m_{\rm{}H}$ is the
proton mass. We take a lower bound value of $1~\kms$ FWHM for
estimating $\sigma_{ic}$. The inter--cloud density $n_{ic}$ is not
well known.  Radiative transfer analysis of CO, \13co, and \c18o\ in
this region (Li 2002) show that the densities are a few times $10^4$
\cc\ in the relatively diffuse area. We use a lower bound of $10^{4}$
\cc\ to calculate the Jeans mass given in Table~\ref{tab1}. The Jeans
mass thus derived will be a conservative upper limit.

The maximum mass which can be supported by a steady $B$ field alone is
\begin{equation}
M_\Phi=c_\Phi\frac{\pi{}Br^2}{G^{1/2}}\lc
\label{mphi}
\end{equation}
where $c_\Phi\sim{}0.12$ is given by numerical simulations
(Mouschovias \& Spitzer 1976; Tomisaka, Ikeuchi \& Nakamura 1988).

Crutcher et al.\ (1999) detect the Zeeman effect in the CN 3mm line
near Orion BN/KL. The field strength is derived to be $0.19~\rm{}mG$
or $0.36~\rm{}mG$, based on different fitting schemes. This is larger
than the $B\sim{}0.03$ mG measured in dark clouds. Given the proximity
of BN/KL to active star formation, the large $B$ could be explained by
rapid collapses' freezing magnetic flux into high density regions
along the line-of-sight.  We thus take a smaller, nominal $B=0.1~\rm{}mG$
for the quiescent Orion cores.

Using the critical mass calculated through Eq.~\ref{mc}, \ref{mj}, and
\ref{mphi}, we define the ``critical mass ratio''
\begin{equation}
R_{c}=\frac{M}{M_{c}}\lp
\label{rc}
\end{equation}
We find that 12 out of 30 (40\%) cores have $R_{c}>1$.

\subsection{Discussion: Is Orion Special?}
Our data suggest mass ratios
$R_{vir}>3$ and $R_c>1$ for most cores. This is different from other
clouds in solar neighborhood within $\sim{}500~\rm{}pc$. In Perseus, Foster et al.\ (2008) and Kirk et al.\ (2007) find an
equivalent $R_{vir}\lesssim{}3$. In Ophiuchus,
Andr{\'e} et al.\ (2007) imply $R_{vir}\le{}1$ based on their Table~4
after adopting the dust opacity used here.  For SCUBA--identified
Gould Belt cores, Sadavoy et al. (2010) find that only 20 out of
354 cores have ${\rm{}M}/{\rm{}M}_J>$ 2 (again adjusting dust opacities, and
excluding their Orion data to avoid overlap). Their M$_J$
represents a purely thermal Jeans mass, which means that these cores
will have an even smaller mass ratio using our definitions.

The observed $R_{vir}$ in Orion may be similar to those found in high
mass star forming regions, e.g., the two clouds studied by Pillai et
al.\ (2011). Although direct comparison is difficult due to the much larger
distance of other high-mass star-forming regions.

The calculations differ slightly between studies, e.g., in the factors
$\alpha$, $\beta$, and dust opacities. These differences alone cannot
explain the larger mass ratios observed in Orion. Thus, our result
that 60\% of Orion cores have $R_{vir}>3$ and 40\% are supercritical
is significant.

In many pixels, the \ammonia\ lines are not well resolved with a FWHM
channel width of 0.62 \kms.  For these pixels, the channel width was
used as the upper limit of the intrinsic line width.  Based on the
combination of the measured line width and line width upper limits,
the velocity dispersion of Orion cores do not exceed those of the
cores in other regions studied by Kirk et al.\ (2007), Andr{\'e} et
al.\ (2007), and Foster et al.\ (2008). The higher values of $R_{vir}$ and $R_c$ of Orion cores are thus
not due to different velocity field, but higher masses. Kauffmann \& Pillai (2010)
suggest that Ophiuchus and Perseus dense cores have a
size--dependent limit
\begin{equation}
m_{\rm{}lim}(r)=870\,M_{\sun}\,(r/{\rm{}pc})^{1.33}\,{},
\label{mlim}
\end{equation}
above which high--mass star formation occurs. We  find 12 cores (40\%) above this threshold.
Within the solar neighborhood, Orion cores seem to be the most likely ones to form high--mass stars.

\section{Conclusions}
We have mapped OMC2 and OMC3 in \ammonia\ (1,1) and
(2,2) with both VLA and GBT.  The combined single dish plus
interferometric data provide a rare detailed look into the thermal and
dynamic properties of a collection of massive quiescent cores. Our
main results are:
\begin{enumerate}
\item We obtain good temperature measurement for 30 dust cores. The
  median core temperature is $17~\rm{}K$. The typical uncertainty for
  derived temperature of each pixel is about $2~\rm{}K$.
\item 12 cores are associated with a protostar.  The average
  temperatures of the protostellar and the starless cores are similar
  suggesting that the heating in OMC2-3 region is primarily external.
\item A total of 24 cores (80\%) are gravitationally bound
  ($R_{vir}>1$), and 11 cores (37\%) achieve $R_{vir}>3$. Compared to
  other Gould Belt clouds, a much higher fraction of cores are tightly bounded.
\item 12 out of 30 cores (40\%) are more massive than the critical mass
  defined as the combination of Jeans mass and mass supported by a
  steady magnetic field of $100~\rm{}\mu{}G$.
\end{enumerate}
In summary, this sample of Orion cores, identified in dust emission
with temperature and turbulence measured in \ammonia\ inversion lines,
are proven to be well bounded by gravity and contains a substantial fraction of
supercritical cores. They will evolve rapidly, either collapsing to form a
star or fragmenting.

 \begin{deluxetable}{lccccccccccccc}
 \tabletypesize{\scriptsize} 
 \tablewidth{0pt}  
 \setlength{\tabcolsep}{0.03in}  
 \tablecolumns{10}  
 \tablecaption{Temperatures and Engergy State of Orion Cores}
\tablehead{  
\colhead{Source} &\colhead{RA}  & \colhead{DEC} & \colhead{$T_{core}$} &\colhead{$D_{core}$}\tablenotemark{a} &\colhead{$\sigma_{core}$}\tablenotemark{b} & \colhead{Mass} & \colhead{M$_J$}& \colhead{M$_\Phi$}  & \colhead{$R_{vir}$} &\colhead{$R_c$} &  \colhead{YSO RA} &\colhead{YSO DEC} & \colhead{Seperation}\tablenotemark{c}\\ 
              & \colhead{(J2000)}&\colhead{(J2000)} &\colhead{K}  &\colhead{Arcsec} &\colhead{\kms} & \colhead{\Ms}& \colhead{\Ms}& \colhead{\Ms}    & \colhead{} & \colhead{} & \colhead{(J2000)}&\colhead{(J2000)} &\colhead{Arcsec}  } 
\startdata 
OriAN-535161-50000&5:35:16&-5:00:00&    16.&     14.&    0.56&   12.&  49.6&   0.6&   2.&  0.2&5:35:16&-5:00:00&   2.7\\
OriAN-535182-50021&5:35:18&-5:00:20&    16.&     16.&    0.48&   26.&  21.9&   0.8&   6.&   1.&5:35:19&-5:00:36&  12.1\\
OriAN-535182-50147&5:35:18&-5:01:46&    21.&     19.&    0.41&    3.&   7.8&   1.2&  0.9&  0.3&5:35:19&-5:01:11&  51.9\\
OriAN-535183-51338&5:35:18&-5:13:37&    31.&     10.&    0.28&  0.3&   0.7&   0.3&  0.7&  0.3&5:35:19&-5:13:11&  35.1\\
OriAN-535189-51412&5:35:18&-5:14:12&    30.&     13.&    0.27&  0.5&   0.7&   0.5&  0.8&  0.4&5:35:21&-5:13:11&  55.4\\
OriAN-535196-51535&5:35:19&-5:15:35&    15.&     21.&    0.47&    6.&  16.3&   1.4&   1.&  0.3&5:35:19&-5:15:36&   2.7\\
OriAN-535207-50053&5:35:20&-5:00:52&    14.&     15.&    0.33&   30.&   3.5&   0.7&  18.&   7.&5:35:19&-5:01:11&  14.2\\
OriAN-535214-51458&5:35:21&-5:14:57&    19.&     18.&    0.37&    4.&   5.5&   1.0&   2.&  0.6&5:35:19&-5:15:00&  26.1\\
OriAN-535216-51039&5:35:21&-5:10:38&    16.&     15.&    0.27&    2.&   0.7&   0.7&   2.&   1.&5:35:23&-5:10:47&  47.1\\
OriAN-535217-51312&5:35:21&-5:13:11&    24.&     18.&    0.52&    9.&  24.1&   1.0&   2.&  0.4&5:35:21&-5:13:11&  18.9\\
OriAN-535218-51422&5:35:21&-5:14:21&    16.&     20.&    0.37&    6.&   5.4&   1.2&   2.&  0.9&5:35:19&-5:15:00&  55.3\\
OriAN-535224-50114&5:35:22&-5:01:14&    19.&     11.&    0.28&   16.&   0.7&   0.4&  28.&  15.&5:35:21&-5:01:11&   0.1\\
OriAN-535225-51014&5:35:22&-5:10:14&    15.&     15.&    0.50&    2.&  29.6&   0.7&  0.5&  0.1&5:35:23&-5:10:47&  36.6\\
OriAN-535229-51240&5:35:22&-5:12:39&    17.&     23.&    0.53&   26.&  20.3&   1.7&   5.&   1.&5:35:23&-5:11:59&  37.4\\
OriAN-535234-51205&5:35:23&-5:12:04&    17.&     18.&    0.36&   13.&   4.4&   1.0&   6.&   2.&5:35:23&-5:11:59&   2.4\\
OriAN-535235-50132&5:35:23&-5:01:32&    27.&     18.&    0.46&   23.&  20.5&   1.0&   5.&   1.&5:35:23&-5:01:11&   3.3\\
OriAN-535235-50734&5:35:23&-5:07:33&    16.&     14.&    0.40&    8.&   9.1&   0.6&   3.&  0.8&5:35:26&-5:07:48&  39.1\\
OriAN-535236-50711&5:35:23&-5:07:10&    17.&     14.&    0.49&    6.&  17.3&   0.6&   2.&  0.3&5:35:26&-5:06:36&  32.4\\
OriAN-535245-50754&5:35:24&-5:07:54&    15.&     16.&    0.42&   12.&   8.2&   0.9&   4.&   1.&5:35:26&-5:07:48&  16.8\\
OriAN-535245-50832&5:35:24&-5:08:31&    18.&     20.&    0.40&   10.&   6.1&   1.3&   3.&   1.&5:35:23&-5:08:23&   3.3\\
OriAN-535255-50237&5:35:25&-5:02:36&    15.&     20.&    0.42&    7.&  13.1&   1.3&   2.&  0.5&5:35:23&-5:01:48&  63.4\\
OriAN-535258-50551&5:35:25&-5:05:50&    19.&     29.&    0.39&   13.&   6.0&   2.6&   3.&   2.&5:35:26&-5:05:59&   7.0\\
OriAN-535261-50126&5:35:26&-5:01:26&    16.&     18.&    0.34&    4.&   0.7&   1.0&   5.&   2.&5:35:23&-5:01:11&  38.9\\
OriAN-535264-50830&5:35:26&-5:08:30&    18.&     15.&    0.51&    4.&  31.4&   0.7&  0.7&  0.1&5:35:26&-5:08:23&  18.9\\
OriAN-535265-50356&5:35:26&-5:03:56&    19.&     32.&    0.68&   13.&  37.6&   3.2&   1.&  0.3&5:35:26&-5:04:12&   1.7\\
OriAN-535265-51011&5:35:26&-5:10:10&    23.&     26.&    0.61&   34.&  36.2&   2.1&   4.&  0.9&5:35:26&-5:10:12&   9.7\\
OriAN-535271-51139&5:35:27&-5:11:39&    20.&     18.&    0.43&    2.&  14.3&   1.0&  0.6&  0.2&5:35:23&-5:11:24&  38.5\\
OriAN-535274-50511&5:35:27&-5:05:11&    13.&     27.&    0.34&   20.&   3.0&   2.3&   7.&   4.&5:35:28&-5:04:47&  15.7\\
OriAN-535276-50935&5:35:27&-5:09:34&    28.&     20.&    0.72&   13.&  88.2&   1.2&   1.&  0.1&5:35:26&-5:09:35&   1.5\\
OriAN-535279-51025&5:35:27&-5:10:25&    17.&     10.&    0.50&    6.&  32.6&   0.3&   2.&  0.2&5:35:26&-5:10:12&  15.5\\
\enddata 
\tablenotetext{a}{$R_{core}$ indicates the geometric mean of the long and short core axes.}
\tablenotetext{b}{$\sigma_{core}$ is derived by averaging all spectra contained in a core and then measuring the line dispersion .}
\tablenotetext{c}{The angular distance between dust contiuum peaks and their respective closest YSOs.}
\label{tab1}
\end{deluxetable} 

 \end{document}